\input{aipcheck}

\documentclass[sort&compress
    ,final            
]
  {aipproc}

\layoutstyle{8x11single}

\usepackage{amssymb}
\usepackage{amsmath}
\usepackage{xspace}                    

\newcommand{\half}{\frac{1}{2}}

\newcommand{\de}{\partial}

\newcommand{\HIGS}{HI$\gamma$S\xspace}
\newcommand{\threeHe}{${}^3$He\xspace}
\newcommand{\fourHe}{${}^4$He\xspace}

\newcommand{\wlab}{\ensuremath{\omega_\mathrm{lab}}}

\newcommand{\alphae}{\ensuremath{\alpha_{E1}}}
\newcommand{\betam}{\ensuremath{\beta_{M1}}}

\newcommand{\gammamm}{\ensuremath{\gamma_{M1M1}}}
\newcommand{\gammaem}{\ensuremath{\gamma_{E1M2}}}
\newcommand{\gammame}{\ensuremath{\gamma_{M1E2}}}

\newcommand{\alphaep}{\ensuremath{\alpha_{E1}^{(\mathrm{p})}}}
\newcommand{\betamp}{\ensuremath{\beta_{M1}^{(\mathrm{p})}}}
\newcommand{\alphaen}{\ensuremath{\alpha_{E1}^{(\mathrm{n})}}}
\newcommand{\betamn}{\ensuremath{\beta_{M1}^{(\mathrm{n})}}}
\newcommand{\alphaes}{\ensuremath{\alpha_{E1}^{(\mathrm{s})}}}
\newcommand{\betams}{\ensuremath{\beta_{M1}^{(\mathrm{s})}}}

\newcommand{\mpi}{\ensuremath{m_\pi}}     
\newcommand{\MeV}{\ensuremath{\mathrm{MeV}}}
\newcommand{\fm}{\ensuremath{\mathrm{fm}}}
\newcommand{\ChiEFT}{$\chi$EFT\xspace}

\newcommand{\NXLO}[1]{N\ensuremath{{}^{#1}}LO\xspace}

\newcommand{\calL}{\mathcal{L}}
\newcommand{\calO}{\mathcal{O}}

\newcommand{\order}[1]{{\cal O}(#1)}

\begin{document}

\title{High-Accuracy Analysis of Compton Scattering in Chiral Effective Field
  Theory: Status and Future}

\classification{
} 

\keywords {Compton scattering, nucleon polarisabilities, spin
  polarisabilities, polarised experiments, Chiral Effective Field Theory,
  $\Delta(1232)$}

\author{\underline{Harald W.~Grie{\ss}hammer}}{%
  address={\mbox{Institute for Nuclear Studies, Department of Physics, George
      Washington University, Washington DC 20052, USA}\\and
    IKP-3, IAS and J\"ulich Centre for Hadron Physics, Forschungszentrum
    J\"ulich, D-52428 J\"ulich, Germany}}

\author{Daniel R.~Phillips}{%
  address={\mbox{Dept.~of Physics and Astronomy, Inst.~of Nuclear and
  Particle Physics, Ohio University, Athens OH 45701, USA}}}

\author{Judith A.~McGovern}{%
  address={School of Physics and Astronomy, The University of Manchester,
  Manchester M13 9PL, UK}}

\begin{abstract}
  Theory and prospects of Compton scattering on nucleons and light nuclei
  below $500\;\MeV$ are outlined; cf.~\cite{Gr12, higherorderpols,
    Griesshammer:2013vga}. Invited contribution at the \textsc{Workshop to
    Explore Physics Opportunities with Intense, Polarized Electron Beams with
    Energy up to 300 MeV}, MIT, Cambridge (USA), 14th-16th March 2013.
\end{abstract}

\maketitle


\section{Why Compton Scattering?}

In Compton scattering $\gamma \text{X}\to\gamma \text{X}$, the electric and
magnetic fields of a real photon induce radiation multipoles by displacing the
charged constituents and currents inside the target. The energy- and
angle-dependence of the emitted radiation explores the interactions of its
constituents. In Hadronic Physics, it elucidates the distribution, symmetries
and dynamics of the charges and currents which constitute the low-energy
degrees of freedom inside the nucleon, and -- for few-nucleon systems -- the
interactions between nucleons; see a recent review for details~\cite{Gr12}.
The 2007 NSAC and 2010 NuPECC Long-Range Plans emphasise therefore the pivotal
r\^ole of the nucleon's temporal two-photon response below $1\,$GeV to
complement the information accessible in one-photon experiments like
form-factor measurements. As a consequence, a number of experiments are
presently being pursued at MAMI (Mainz)~\cite{Miskimen}, HI$\gamma$S at
TUNL~\cite{We09}, and MAX-Lab at Lund~\cite{Fe08}. For the next generation,
Aron Bernstein and Rory Miskimen have shown at this workshop how to use
high-intensity electron beams like MESA for ``near-real'' photon
experiments~\cite{Bernstein, Miskimen}.

In contradistinction to many electro-magnetic processes, such structure
effects have only recently been subjected to a multipole-analysis.  The
Fourier transforms of the corresponding temporal response functions are the
proportionality constants between incident field and induced multipole. These
\emph{energy-dependent polarisabilities} parametrise the stiffness of the
nucleon $N$ (spin $\frac{\vec{\sigma}}{2}$) against transitions ${Xl\to
  Yl^\prime}$ of given photon multipolarity at fixed frequency $\omega$
($l^\prime=l\pm\{0;1\}$; $X,Y=E,M$; $T_{ij}=\half (\de_iT_j + \de_jT_i)$;
$T=E,B$). Up to $500\;\MeV$, the relevant terms are:
\begin{equation}
\begin{split}
  \label{polsfromints}
  \calL_\text{pol}=2\pi\;N^\dagger
  \;\big[&{\alpha_{E1}(\omega)}\;\vec{E}^2\;+
  \;{\beta_{M1}(\omega)}\;\vec{B}^2\; 
 +\;{\gamma_{E1E1}(\omega)}
  \;\vec{\sigma}\cdot(\vec{E}\times\dot{\vec{E}})\;
  +\;{\gamma_{M1M1}(\omega)}
  \;\vec{\sigma}\cdot(\vec{B}\times\dot{\vec{B}})
  \\&
  -\;2{\gamma_{M1E2}(\omega)}\;\sigma^i\;B^j\;E_{ij}\;+
  \;2{\gamma_{E1M2}(\omega)}\;\sigma^i\;E^j\;B_{ij} \;+\;\dots\;(\mbox{photon
    multipoles beyond dipole}) \big]\;N
\end{split}
\end{equation} 
The two spin-independent polarisabilities $\alpha_{E1}(\omega)$ and
$\beta_{M1}(\omega)$ parametrise electric and magnetic dipole transitions.
Of particular interest are at present the four dipole spin-polarisabilities
$\gamma_{E1E1}(\omega)$, $\gamma_{M1M1}(\omega)$, $\gamma_{E1M2}(\omega)$,
$\gamma_{M1E2}(\omega)$. They encode the response of the nucleon
spin-structure, i.e.~of the spin constituents. Intuitively interpreted, the
electromagnetic field associated with the spin degrees causes bi-refringence
in the nucleon, just like in the classical Faraday-effect. Only the linear
combinations $\gamma_0$ and $\gamma_\pi$ of scattering under $0^\circ$ and
$180^\circ$ are somewhat constrained by data or phenomenology, with
conflicting results for the proton (MAMI, LEGS) and large error-bars for the
neutron. 

The values $\alpha_{E1}(\omega=0)$ etc. are often called ``the (static)
polarisabilities''; but the $\omega$-dependence reveals more.
Since the polarisabilities are the parameters of a multipole decomposition,
they contain not more information than the full amplitudes, but characteristic
signatures in specific channels are easier to interpret. For example, the
significant $\omega$-dependence of $\beta_{M1}(\omega)$ and
$\gamma_{M1M1}(\omega)$ already for $\omega\gtrsim100\;\MeV$ comes from the
strong para-magnetic $\gamma \text{N}\Delta(1232)$ transition. The
$\Delta(1232)$ enters thus dynamically well below the resonance region. The
electric polarisabilities exhibit a pronounced cusp at the pion-production
threshold. As soon as an inelastic channel opens, polarisabilities become
complex. Thus, their imaginary parts provide an alternative to explore
pion-photoproduction multipoles.  Polarisabilities also enter as one of the
bigger sources of uncertainties in theoretical determinations of the
proton-neutron mass difference (see e.g.~most recently
\cite{WalkerLoud:2012bg}), and of the two-photon-exchange contribution to the
Lamb shift in muonic hydrogen (see e.g.~most recently \cite{Bi12}).
Finally, nuclear targets provide an opportunity to study not only neutron
polarisabilities, but also the nuclear force directly, since the photons
couple to the charged pion-exchange currents in the nucleus.

\section{Chiral Effective Field Theory \ChiEFT in Compton Scattering}

Interpreting such data of course requires commensurate theoretical support. One
must carefully evaluate data-consistency in one model-independent framework
for hidden systematic errors; subtract binding effects in few-nucleon systems;
extract the polarisabilities using minimal theoretical bias; identify the
underlying QCD mechanisms, like the detailed chiral dynamics of the pion cloud
and of the $\Delta(1232)$ resonance; relate them to emerging lattice QCD
simulations -- and do all of that while providing reproducible theoretical
uncertainties.

\ChiEFT, the low-energy theory of QCD and extension of Chiral Perturbation
Theory to few-nucleon systems, has been quite successful in proton and
few-nucleon Compton scattering, starting with the parameter-free leading-order
prediction $\alphae=10\betam=12.4\times10^{-4}\;\fm^3$ by Bernard et
al.~\cite{Be91}. \ChiEFT generates the most general Compton amplitude
consistent with gauge invariance, the pattern of chiral-symmetry breaking in
QCD, and Lorentz covariance. A particularly interesting \ChiEFT prediction is
that small proton-neutron differences in polarisabilities stem from
chiral-symmetry breaking $\pi$N interactions and thus probe details of QCD.
In \ChiEFT with explicit $\Delta(1232)$ degrees of freedom, the low-energy
scales are the pion mass $\mpi\approx 140\;\MeV$ as the typical chiral scale;
the Delta-nucleon mass splitting $\Delta_M\approx290\;\MeV$; and the photon
energy $\omega$.  When measured in units of a natural ``high'' scale
$\Lambda\gg\Delta_M,\mpi,\omega\approx800\;\MeV$ at which this variant can be
expected to break down because new degrees of freedom become dynamical, each
gives rise to a small, dimensionless expansion parameter. In the
$\delta$-expansion of Pascalutsa and Phillips~\cite{PP03}, one avoids a
threefold expansion by approximately relating scales so that only one
dimensionless parameter is left: 
\begin{equation}
  \delta\equiv
  \frac{\Delta_M}{\Lambda}\approx\left(\frac{m_\pi}{\Lambda}\right)^{1/2}
  \;\;,
  \label{eq:delta}
\end{equation}
i.e.~numerically $\delta\approx0.4$. For $\omega\sim\mpi$, the Thomson
amplitude is leading-order, $\order{e^2\delta^0}$; structure effects start
with $\pi$N loops at $\order{e^2\delta^2}$; and since $\Delta_M \sim \delta$
whereas $m_\pi \sim \delta^2$, $\pi \Delta$ loops are suppressed by an
additional power to $\order{e^2 \delta^3}$.

The Delta-pole graph is $\order{e^2 \delta^3}$ for $\omega \sim m_\pi$, but
its enhancement close to the Delta's on-shell point leads to a re-ordering of
contributions at higher energies, $\omega \sim \Delta_M$.  The $\pi \text{N}$
loops that generate the resonance's nonzero width must then be resummed. In
this r\'egime, the dominant Compton mechanism, $\order{e^2 \delta^{-1}}$, is
the excitation of a dressed $\Delta(1232)$ by the magnetic transition from the
nucleon, followed by de-excitation via the same M1 transition. The E2
$\text{N} \rightarrow \Delta(1232)$ transition and leading-one-loop
corrections to the $\gamma \text{N}\Delta$ vertex enter at $\order{e^2
  \delta^0}$. Relativistic kinematics is of course essential around the
$\Delta$ resonance. Recently, single-nucleon Compton amplitudes were derived
which apply from zero energy to about $400\;\MeV$. For $\omega\lesssim\mpi$,
they contain all contributions at $\calO(e^2\delta^4)$ (\NXLO{4}, accuracy
$\delta^5\lesssim2\%$), and for $\omega\sim \Delta_M$ all at
$\calO(e^2\delta^0)$ (NLO, accuracy $\delta^2\lesssim20\%$); see
Refs.~\cite{Gr12, higherorderpols} for a detailed discussion. Vladimir
Pascalutsa's talk presented results of an alternative, manifestly covariant
\ChiEFT variant~\cite{Pascalutsa}.

For light nuclei, $\chi$EFT provides at present deuteron Compton scattering
results which are complete at $\calO(e^2\delta^3)$ from the Thomson limit up
to about $120$~MeV, including the $\Delta(1232)$ degree of freedom~\cite{Hi05,
  Gr12}. This is now being extended both to higher energies and by including
the new single-nucleon amplitudes. For \threeHe, the results in a variant
without explicit $\Delta(1232)$ and for
$\omega\in[80;120]\;\MeV$~\cite{Shukla:2008zc} are extended, too. In
few-nucleon systems, the Thomson limit as exact low-energy theorem is a result
of $\text{N}\text{N}$ rescattering between photon emission and absorption.
While computationally intensive, its r\^ole diminishes for
$\omega\gtrsim100\;\MeV$ because the photon does not scatter any more
coherently from the target nucleus as a whole. Finally, nuclei themselves are
made of charged particles which are displaced by the photon fields and thus
have an intrinsic polarisability. For the deuteron, these are known on the
$\lesssim1\%$-level, with various EFT variants and conventional calculations
agreeing very well~\cite{Gr12}.

\section{New Static Polarisabilities from $\chi$EFT}

In the unified single-nucleon amplitudes described above, the $\pi$N
parameters take their standard values~\cite{Gr12, higherorderpols}. The
$\Delta(1232)$ parameters $\Delta_M= 293$~MeV and $g_{\pi \text{N}\Delta}
=1.425$ are obtained from the Breit-Wigner peak and width via the relativistic
formula, and the ratio of E2 and M1 couplings is $b_2/b_1=-0.34$. Two contact
interactions encode the short-distance ($r \ll 1/m_\pi,1/\Delta_M$)
contributions to the scalar polarisabilities.  Their coefficients (or,
equivalently, the static values $\alphae$ and $\betam$) must be fitted. To
that end, one must first establish a statistically consistent database from
all available proton data below $350\;\MeV$ in Ref.~\cite{Gr12}, carefully
pruning the data by objective and transparent criteria.

Since the power counting confirms that the high-energy amplitudes are most
sensitive to $\Delta$ parameters, the $\gamma$N$\Delta$ M1 coupling $b_1$ is
determined from the MAMI data for $\wlab=$200--325~MeV. Sensitivity to the
polarisabilities is greater at $\omega\lesssim170$~MeV, where the amplitudes
are also known with higher accuracy. Thus, $\alphaep$ and $\betamp$ are fit
concurrently to these low-energy data, with iteration betwixt both regions
until convergence is
reached. 
One finds a solution with a $\chi^2/{\rm d.o.f.}= 113.2/135$, $b_1=3.61\pm
0.02$ and the static scalar proton polarisabilities as (stat.~errors from
$\chi^2+1$)~\cite{higherorderpols}:
\begin{eqnarray}
  \alphaep
  =
  [10.7\pm0.4_\mathrm{stat}\pm0.2_\mathrm{Baldin}\pm0.3_\mathrm{theory}]
  \times10^{-4}\;\fm^3
  &\!\!\!\!\!\!,\!\!\!\!& 
  \betamp 
  =
  [3.1\mp0.4_\mathrm{stat}\pm0.2_\mathrm{Baldin}\mp0.3_\mathrm{theory}]
  \times10^{-4}\;\fm^3\;\;\;\;
\end{eqnarray}
Since a fit to $\alphae$ and $\betam$ independently is highly consistent with
the Baldin sum rule $\alphaep+\betamp = [13.8\pm0.4] \times10^{-4}\;\fm^3$,
the numbers quoted above use this constraint. All fits are stable under
reasonable variations in the procedure and agree with the data
well beyond the region in which the parameters are determined.
Special care has been taken to reproducibly justify a theoretical
uncertainty of $\pm0.3\times10^{-4}\;\fm^3$ from the most
conservative of several estimates of higher-order terms. 
For an acceptable fit, $\gammamm$ is treated as parameter, albeit its counter
term enters at higher order.

Neutron polarisabilities are extracted from the elastic deuteron data base of
Ref.~\cite{Gr12}. It has significantly larger statistical error-bars and is
only a tenth of the size of the proton one, with data at only a few angles and
energies $\omega\in[49;94]\;\MeV$. These amplitudes are one order lower than
in the proton extraction, but the statistical errors are still larger than the
estimated theoretical
uncertainties. 
The correct Thomson limit and $\text{N}\text{N}$ rescattering is important
also to reduce residual dependence 
on the choice of the deuteron wave function to $< 1$\%~\cite{Gr12}.
The fit to the isoscalar, scalar dipole polarisabilities yields with
$\chi^2/{\rm d.o.f.}=24.3/25$: 
\begin{eqnarray}
  \alphaes =
  [10.9\pm0.9_\mathrm{stat}\pm0.2_\mathrm{Baldin}\pm0.8_\mathrm{theory}]
  \times10^{-4}\;\fm^3
  &\!\!\!\!\!\!,\!\!\!\!&
  \betams=
  [3.6\mp0.9_\mathrm{stat}\pm0.2_\mathrm{Baldin}\mp0.8_\mathrm{theory}]
  \times10^{-4}\;\fm^3\;\;\;\;\\
\alphaen=
[11.1\pm 1.8_\mathrm{stat}\pm0.2_\mathrm{Baldin}\pm0.8_\mathrm{theory}]
  \times10^{-4}\;\fm^3
 &\!\!\!\!\!\!,\!\!\!\!&
\betamn=
[4.2\mp 1.8_\mathrm{stat}\pm0.2_\mathrm{Baldin}\mp0.8_\mathrm{theory}]
  \times10^{-4}\;\fm^3\;\;\;\;
\end{eqnarray} 
The $\chi^2+1$ ellipsoid is again the statistical error. In the last
line, the proton and isoscalar values are combined to the static scalar
neutron polarisabilities. An independent fit to $\alphaes$ and $\betams$ is
again consistent with the (isoscalar) Baldin sum rule, so this constraint
reduces statistical uncertainties. In contrast to the proton case, the
data are consistent: each experiment contributes roughly equally to the
$\chi^2$, and the extracted polarisabilities are largely insensitive to the
elimination of any one data set.
Within the statistics-dominated errors, the proton and neutron
polarisabilities are thus identical, i.e.~no isospin breaking effects of the
pion cloud are seen, as predicted by Chiral EFT.  In all cases, the
normalisation of each data set is floated within the quoted normalisation
uncertainty; Refs.~\cite{Gr12,higherorderpols} give more details.

\begin{figure}[!htb]
 \parbox{0.36\linewidth}{%
   \includegraphics*[width=\linewidth]{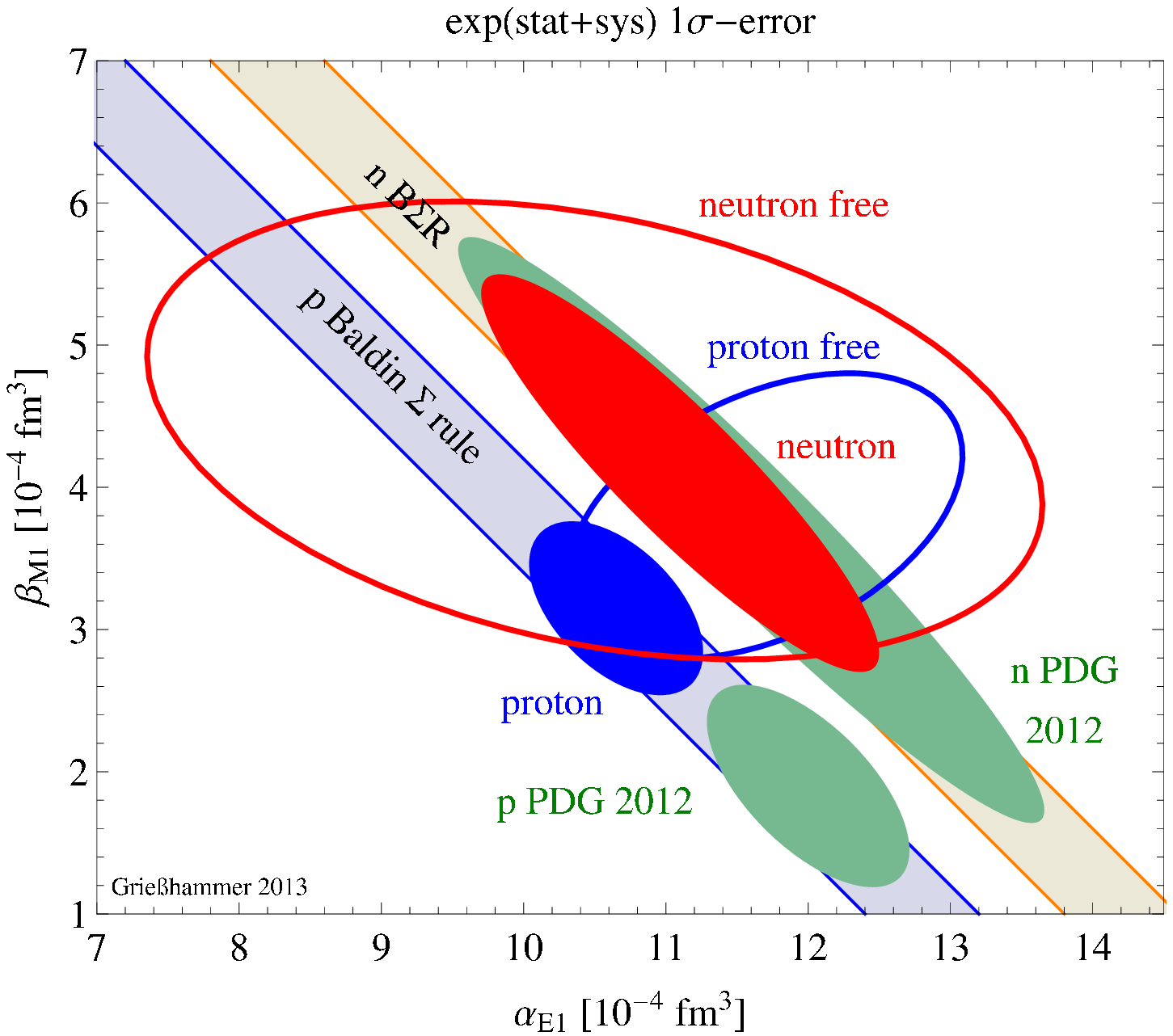}}
\hspace*{3ex}
 \parbox{0.6\linewidth}{%
   \includegraphics*[height=\linewidth,angle=-90]{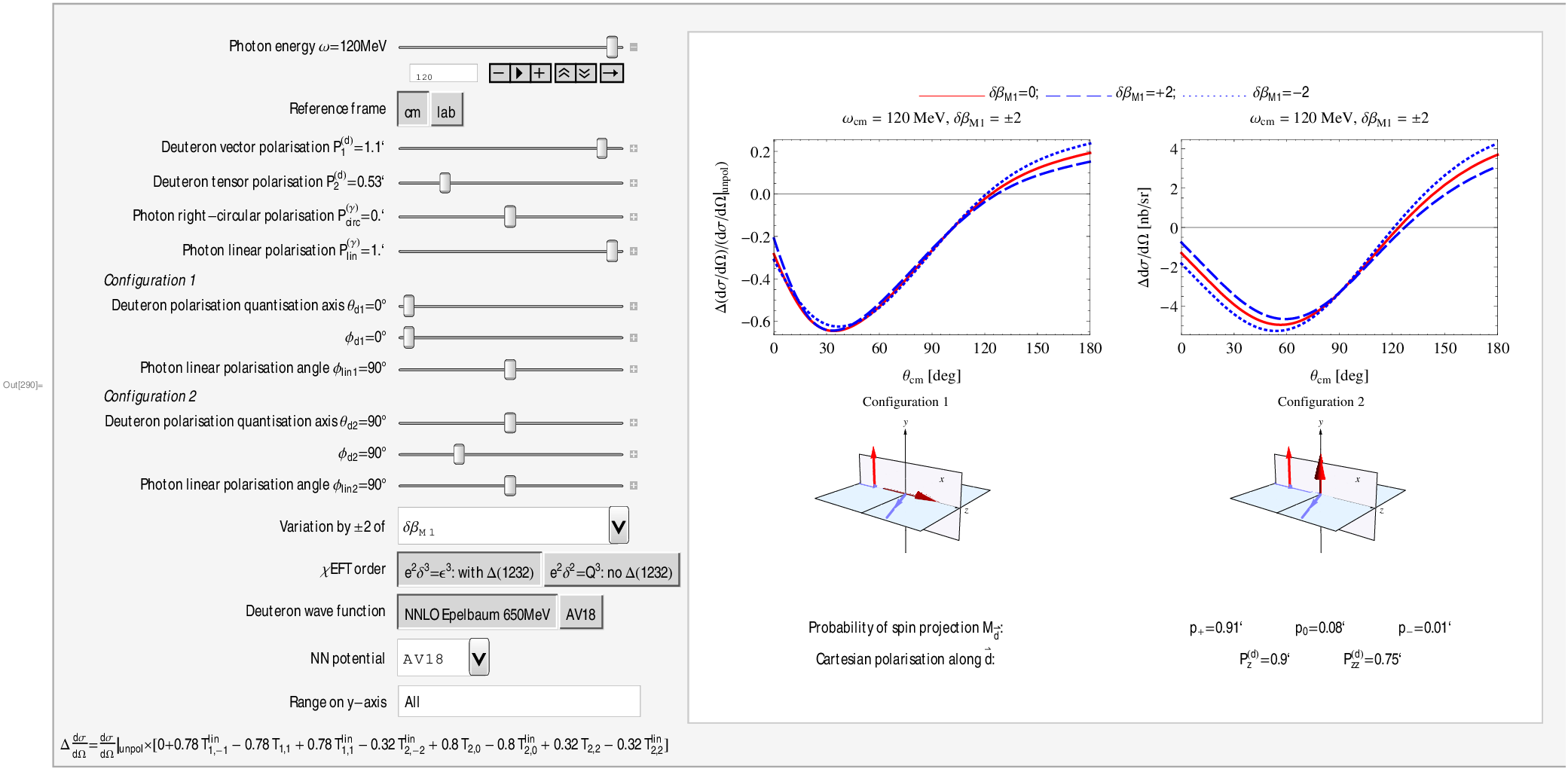}}
 \caption{Left: Comparison of static scalar dipole polarisability values in
   \ChiEFT from Refs.~\cite{Gr12, higherorderpols} and PDG values (green).
   Blue: proton; red: neutron; disks (ellipses): with (without) Baldin Sum
   Rule constraints. Notice that errors are statistical only, and $1\sigma$,
   \emph{not} $\chi^2+1$. Right: Screen-shot of a \emph{Mathematica} notebook
   for $\gamma$d scattering with arbitrary polarisations; from
   Ref.~\cite{Griesshammer:2013vga}.
  \label{fig:screenshot}}
\end{figure}

\section{Opportunities for High-Accuracy, High-Luminosity Experiments}

The future lies in un-, single- and double-polarised experiments of high
accuracy, with reproducible systematic uncertainties; see~\cite{Miskimen,
  We09, Fe08} for ongoing and planned efforts. For unpolarised experiments on
the proton, the greatest impact is where the data base of nearly $300$ points
is most scarce: (1) at forward angles (test the Baldin sum rule); (2) at
extreme back-angles (test $\alphae-\betam$); (3) between $170$ and $240\;\MeV$
(sparse and inconsistent data); (4) to resolve the several-$\sigma$
discrepancy between LEGS and MAMI which prevents a consistent fit to all sets
simultaneously~\cite{higherorderpols}.

In order to understand the subtle differences of the pion clouds around the
proton and neutron induced by explicit chiral symmetry breaking in QCD, we
need the neutron polarisabilities with uncertainties comparable to those of
the proton. MAXlab and \HIGS are aiming to augment the angular and energy
range of the $29$ unpolarised data points for the deuteron with statistical
and systematic uncertainties of better than $5\%$. Both also have plans for
\threeHe. First results on unpolarised ${}^6$Li have been reported, and theory
is starting to catch up~\cite{Bampa:2011fq}. Such targets are experimentally
better to handle and provide count rates which scale at least linearly with
the target charge when photons scatter incoherently from the protons,
i.e.~$\omega\gtrsim100\;\MeV$.  However, describing the energy levels of these
nuclei with adequate accuracy is theoretically involved. For the proton,
amplitudes on the $\lesssim2\%$-level are available; for deuteron and
\threeHe, consistent Compton amplitudes from zero energy into the Delta
resonance region are being developed in \ChiEFT. Around
\threeHe-\fourHe-${}^6$Li may well be the ``sweet-spot'' between needs and
wants of theorists and experimentalists.  Neutron values can also be
isolated both in quasi-free kinematics and from polarised \threeHe, which
effectively is a free-neutron target~\cite{Shukla:2008zc}.

The highest impact of high-intensity (near-real) photon beam machines will
however be in the study of the so-far nearly untested spin-polarisabilities:
four each for the proton and neutron. Since they test the spin-constituents of
the nucleon, they are a top priority of experiment and theory alike.
Sensitivity studies have been performed in \ChiEFT variants with and without
explicit $\Delta(1232)$; see summary in~\cite[Sec.~6.1]{Gr12} and
V.~Pascalutsa's talk for a covariant \ChiEFT variant~\cite{Pascalutsa}.
Asymmetries remove many systematic effects. Recently, the cross section with
arbitrary photon and deuteron polarisation has for example been parametrised
via 18 independent observables~\cite{Griesshammer:2013vga}.  An exploration of
the sensitivity of each on the nucleon's scalar and spin dipole
polarisabilities in the \ChiEFT variant discussed above shows that some
asymmetries are sensitive to only one or two dipole polarisabilities. This
makes them particularly attractive for an energy-dependent multipole-expansion
of Compton scattering; cf.~\cite{Gr12}. For spin polarisabilities with an
error of $\pm2\times10^{-4}\;\fm^4$, asymmetries should be measured with an
accuracy of $\gtrsim10^{-2}$, with differential cross sections of a dozen
nb/sr at $100\;\MeV$ or a few dozen nb/sr at $250\;\MeV$.  Relative to
single-nucleon Compton scattering, interference with the $D$ wave and the
pion-exchange current of the deuteron increases the sensitivity to the
``mixed'' spin polarisabilities $\gammaem$ and $\gammame$. One may thus
speculate that their determination will first appear from deuteron data -- and
high-intensity beams can provide the necessary accuracy.  A \emph{Mathematica
  9.0} file for $\omega<120\;\MeV$ is available from \texttt{hgrie@gwu.edu};
see screen-shot in Fig.~\ref{fig:screenshot}. Parallel studies for the proton
and \threeHe are under way.

\begin{theacknowledgments}
  HWG cordially thanks the organisers for a stimulating atmosphere and
  financial support.
  Work supported in part by UK Science and Technology Facilities Council
  grants ST/F012047/1, ST/J000159/1 (JMcG) and ST/F006861/1 (DRP), by US
  Department of Energy grants DE-FG02-95ER-40907 (HWG) and DE-FG02-93ER-40756
  (DRP)
, by the Deutsche Forschungsgemeinschaft and the National Natural Science
  Foundation of China via the Sino-German CRC 110
  ``Symmetries and the Emergence of Structure in QCD'' (HWG), and by the EPOS
  network of the European Community Research Infrastructure Integrating
  Activity ``Study of Strongly Interacting Matter'' (HadronPhysics3).
\end{theacknowledgments}


%
\bibliographystyle{aipproc}   
\end{document}